\documentclass{elsart}
\usepackage{amssymb,graphicx}

\begin{document}
\journal{Physics Letters A}
 
\begin{frontmatter}
 
\title{Superiority of semiclassical over quantum mechanical calculations
       for a three-dimensional system}
\author{J\"org Main and G\"unter Wunner}
\address{Institut f\"ur Theoretische Physik 1, Universit\"at Stuttgart,
	 D-70550 Stuttgart, Germany}
\author{Erdin\c c At\i lgan and Howard S.\ Taylor}
\address{Department of Chemistry, University of Southern California,
         Los Angeles, California 90089}
\author{Paul A.\ Dando}
\address{Department of Physics and Astronomy,
         University College London, Gower Street, London WC1E 6BT, UK}
\maketitle

\begin{abstract}
In systems with few degrees of freedom modern quantum calculations are,
in general, numerically more efficient than semiclassical methods.
However, this situation can be reversed with increasing dimension of 
the problem.
For a three-dimensional system, viz.\ the hyperbolic four-sphere scattering
system, we demonstrate the superiority of semiclassical versus quantum 
calculations.
Semiclassical resonances can easily be obtained even in energy regions which
are unattainable with the currently available quantum techniques.

\end{abstract}

\end{frontmatter}

\section{Introduction}
The numerical calculation of quantum spectra is a nontrivial task
especially for nonintegrable systems with many degrees of freedom.
Rigorous computational methods have been developed for directly solving
Schr\"odinger's equation, e.g., by time-dependent wave packet expansions or
numerical diagonalization of the Hamiltonian in a complete basis set.
Exact quantum mechanical calculations usually require storage of 
multidimensional wave functions and a computational effort that grows
exponentially with the number of coupled degrees of freedom.
These methods are therefore feasible for systems with relatively few
degrees of freedom.
As an alternative to exact quantum calculations, approximate, e.g.\
semiclassical, methods can be applied.
For systems with a chaotic classical dynamics Gutzwiller's trace formula 
expresses the quantum density of states in terms of the periodic orbits of 
the underlying classical system \cite{Gut90}, i.e.,
\begin{equation}
 \varrho(E) = \varrho_0(E) - \frac{1}{\pi}\, {\rm Im}\,
   \sum_{\rm po} {A}_{\rm po} {\rm e}^{{\rm i}S_{\rm po}/\hbar} \; ,
\label{rho_sc}
\end{equation}
where $\varrho_0(E)$ is the mean density of states given by the phase space
volume, and ${A}_{\rm po}$ and $S_{\rm po}$ are the amplitudes
(including a phase given by the Maslov index) and classical actions of the 
classical periodic orbits (po), respectively.
Eq.~(\ref{rho_sc}) can be applied to systems with an arbitrary number of
degrees of freedom.
However, the number of periodic orbits and the numerical effort needed to 
find them usually increases very rapidly with increasing dimension of the 
phase space.
As a matter of fact, Gutzwiller's periodic orbit theory has been applied 
predominantly to systems with two degrees of freedom, e.g., the anisotropic 
Kepler problem \cite{Gut90,Tan91}, the hydrogen atom in a magnetic 
field \cite{Tan96}, and two-dimensional billiards \cite{Cvi89,Eck95,Kea94}.
For these systems direct quantum mechanical computations are usually
more powerful and efficient than the semiclassical calculation of spectra
by means of periodic orbit theory.
Practical applications of periodic orbit theory to three-dimensional systems
are very rare.
For the three-dimensional Sinai billiard extensive quantum computations have
been performed and the quantum spectra have been analyzed in terms of 
classical periodic orbits \cite{Pri95,Pri00}.
However, no semiclassical eigenstates have been calculated from the set of
periodic orbits.
Semiclassical resonances have been obtained for the three-dimensional 
two- and three-sphere scattering systems \cite{Hen97} but for these systems
all periodic orbits lie in a one- or two-dimensional subspace.

In this Letter we will, for the first time, calculate semiclassical resonances
for a billiard system with genuinely three-dimensional periodic orbits, viz.\ 
the scattering of a particle on four equal spheres centered at the corners 
of a regular tetrahedron.
Recently, experiments on chaotic light-scattering from the four-sphere system 
have attracted much attention \cite{Swe99,Swe01}.
An exact quantum mechanical recipe for the computation of resonances has
been introduced in Ref.~\cite{Hen97}.
We will demonstrate that for this system semiclassical methods are superior
to direct quantum mechanical computations, i.e., semiclassical resonances 
can easily be obtained even in energy regions which are unattainable with 
the presently known quantum techniques.

\section{The four-sphere scattering system}
The four-sphere system can be regarded as the natural genuinely 
three-dimen\-sional generalization of the two-dimensional three-disk system,
which has served as the prototype model for classical, semiclassical, and
quantum investigations of a chaotic repellor \cite{Gas89},
as well as the development of cycle-expansion methods \cite{Cvi89,Eck95}.
The periodic orbits of the symmetry reduced three-disk system can be
described by a binary symbolic code, and allow us to compute the
semiclassical resonances when Gutzwiller's trace formula (\ref{rho_sc}) is 
combined, e.g., with the cycle-expansion \cite{Cvi89,Eck95} or harmonic 
inversion \cite{Mai97c,Mai99} technique.
The semiclassical resonances are approximations to the exact quantum 
mechanical ones, which are obtained, e.g., in the $A_1$ subspace, 
as zeros of the determinant of the matrix \cite{Gas89,Wir99}
\begin{eqnarray}
 {\mathbf M}(k)_{mm'} = \delta_{mm'}
   &+& d_m d_{m'}\frac{J_m(ka)}{H_{m'}^{(1)}(ka)}
   \left\{\cos\left(\frac{\pi}{6}(5m-m')\right)H_{m-m'}^{(1)}(kR) 
    \right. \nonumber \\
         &+& (-1)^{m'} \left. \cos\left(\frac{\pi}{6}(5m+m')\right)
   H_{m+m'}^{(1)}(kR) \right\}
\label{M_3disk}
\end{eqnarray}
with $0\le m, m' < \infty$ and
\[
 d_m = \left\{
  \begin{array}{r}\sqrt{2} \mbox{~for~} m>0\\
               1\mbox{~for~} m=0
  \end{array}\right. \; .
\]
In Eq.~(\ref{M_3disk}) $a$ and $R$ are the radius and center-to-center 
separation of the disks, $k$ is the (complex) wave number, and $J_m(x)$ and 
$H_m^{(1)}(x)$ are Bessel and Hankel functions.
The size of the matrix ${\mathbf M}(k)_{mm'}$ can be truncated by an upper 
angular momentum $m_{\rm max}\gtrsim 1.5ka$ \cite{Wir99}.
For the frequently chosen disk separation $R=6a$ both the semiclassical and
quantum mechanical methods allow for the efficient computation of resonances
in the region $0\le {\rm Re}~ka \le 250$.

In the three-dimensional four-sphere scattering system the computation of both
the semiclassical and quantum mechanical resonances becomes more expensive.
However, as will be shown below, the numerical effort required for the quantum 
calculations increases much more rapidly than that for the semiclassical.
For identical spheres with radius $a$ and equal separation $R$ the discrete
symmetry of the tetrahedral group, $T_d$, reduces the spectroscopy to five
irreducible subspaces: $A_1$, $A_2$, $T_1$, $T_2$, and $E$ \cite{Ham62}.
We will perform all calculations for the $A_1$ subspace in what follows.

Exact quantum resonances of the four-sphere system can be obtained as roots
of the equation
\begin{equation}
 \det {\mathbf M}(k)_{lm,l'm'} = 0 \; ,
\label{M_4sphere}
\end{equation}
with $0\le l,l'\le l_{\rm max}$ and $m,m'=0,3,6,9,\dots,l_{\rm max}$.
The truncation value $l_{\rm max}$ for the angular momentum can be estimated
by $l_{\rm max} \gtrsim 1.5ka$.
The calculation of the matrix elements ${\mathbf M}(k)_{lm,l'm'}$ is rather
complicated, as compared to the simple result in Eq.~(\ref{M_3disk}).
Explicit expressions are given in Ref.~\cite{Hen97}.
(Note that $g_{m=0}$ should read $g_0=1/\sqrt{2}$ instead of $\sqrt{2}$ in 
Eq.~(38) of \cite{Hen97}.)
However, the serious problem of solving Eq.~(\ref{M_4sphere}) is the 
scaling of the dimension of the matrix ${\mathbf M}(k)_{lm,l'm'}$, which is
an $N\times N$ matrix with $N=(l_{\rm max}+2)(l_{\rm max}+3)/6$, i.e.,
$N$ scales as $N\sim k^2$ for the four-sphere system, as compared to
$N\sim k$ for the three-disk system, Eq.~(\ref{M_3disk}).
For example, in the region $ka\approx 200$ the required matrix dimension is 
$N\gtrsim 300$ for the three-disk compared to $N\gtrsim 15000$ for the 
four-sphere system.
With currently available computer technology it is, therefore, impossible
to significantly extend the quantum calculations for the four-sphere system 
to the region $ka\gg 50$ using the method of Ref.~\cite{Hen97}. 

The computation of resonances in the region, e.g., $ka\le 250$ is, however,
not a problem when periodic orbit theory is used.
Eq.~(\ref{rho_sc}) is valid for the four-sphere system with the 
periodic orbit sum now including all three-dimensional periodic orbits
which are scattered between the four spheres.
In full coordinate space each periodic orbit can be described by a symbolic
code given by the sequence of spheres where the orbit is scattered.
Due to the $T_d$ symmetry of the problem each orbit can be symmetry reduced
to the fundamental domain.
The symmetry reduced orbits can be described by a ternary alphabet of symbols
`0', `1', and `2', which are the three fundamental orbits, i.e., the
symmetry reductions of the shortest orbits scattered between two, three, and
four spheres, respectively.
We shall use the symbol `0' for returning back to the previous sphere after 
one reflection, symbol `1' for the reflection to the other third sphere
out of the incident direction but in the same reflection plane of the orbit,
and symbol `2' for the reflection to the other fourth sphere out of the 
reflection plane of the orbit.
The reflection plane is defined as the plane which contains the centers 
of the first three different spheres backward in the history of the 
itinerary code of the orbit.
Note that orbit codes which contain only the symbols `0' and `1' lie in a 
two-dimensional plane, i.e., they correspond to the set of orbits with a 
binary symbolic code, which 
has been well-established for the three-disk system \cite{Cvi89,Eck95}.
Orbits including the `2'-symbol are genuinely three-dimensional orbits.
The periodic orbits can be invariant under certain rotations and reflections.
Each orbit can be assigned one of the symmetry elements
$\{e,\sigma_d,C_2,C_3,S_4\}$ of the group $T_d$.
Symmetry reduced orbits in the fundamental domain are two-, three-,
or four-times shorter than the orbit in the full coordinate space when
they belong to the symmetry class $\{\sigma_d,C_2\}$, $C_3$, or $S_4$, 
respectively.
The length of orbits belonging to symmetry class $e$ is unchanged under 
symmetry reduction.

For the numerical calculation of the periodic orbits we vary, for a given
symbolic code, the reflection points on the spheres until the physical length 
of the orbit becomes a minimum.
Note that for an orbit with symbol length $n$ the number of variational
parameters is $2n$ for the four-sphere system, as compared to $n$ parameters
for the three-disk system.
Despite this, the periodic orbit search remains numerically very efficient.
In chaotic systems the number of periodic orbits increases exponentially.
For the three-disk system the number of orbits with cycle length $n$ is
given approximately by $N\sim 2^n/n$ whereas it scales as $N\sim 3^n/n$
for the four-sphere system.
Nevertheless, orbits up to a sufficiently high cycle length can be obtained.
For a sphere separation of $R=6a$ we have calculated the complete set of 
primitive periodic orbits with cycle length $n\le 14$, numbering 533830 
orbits in total.
More details about the symbolic code, the symmetry properties, and the
numerical computation of the periodic orbits will be given elsewhere.

The calculation of the periodic orbit amplitudes ${A}_{\rm po}$ in
(\ref{rho_sc}) requires the knowledge of the monodromy matrices and the
Maslov indices of the orbits.
The Maslov index increases by 2 at each reflection on a hard sphere, i.e.,
$\mu_{\rm po}=2n$ for an orbit with cycle length $n$.
The calculation of the monodromy matrix ${\mathbf M}_{\rm po}$ for the 
periodic orbits of three-dimensional billiards has been investigated in 
Refs.~\cite{Pri00,Sie98}.
${\mathbf M}_{\rm po}$ is a symplectic $(4\times 4)$ matrix with eigenvalues
$\lambda_1$, $1/\lambda_1$, $\lambda_2$, and $1/\lambda_2$.
For the hyperbolic four-sphere system $\lambda_1$ and $\lambda_2$ are either
both real or the orbits are loxodromic, i.e., the eigenvalues of
${\mathbf M}_{\rm po}$ are a quadruple 
$\{\lambda, 1/\lambda, \lambda^\ast, 1/\lambda^\ast\}$ with $\lambda$ being
a complex number.
The periodic orbit sum for the four-sphere system then reads
\begin{equation}
 g(k) = \sum_p \sum_{r=1}^\infty
   \frac{(-1)^{rn_p}L_p{\rm e}^{{\rm i}krL_p}}
        {\sqrt{|(2-\lambda_{p,1}^r-\lambda_{p,1}^{-r})
                (2-\lambda_{p,2}^r-\lambda_{p,2}^{-r})|}} \; ,
\label{gk}
\end{equation}
where $n_p$ is the cycle length, $L_p$ the physical length, $\lambda_{p,i}$
are the eigenvalues of the monodromy matrix, and $r$ is the repetition number
of the primitive periodic orbit $p$.
The parameters for the primitive periodic orbits with cycle length $n_p\le 3$
are given in Table~\ref{table1}.
\begin{table}[t]
\caption{\label{table1}
Parameters of the symmetry reduced primitive periodic orbits $p$ with 
cycle length $n_p\le 3$ of the four-sphere system with radius $a=1$ and 
center-to-center separation $R=6$.}
\begin{center}
\begin{tabular}[t]{lcrrrrr}
  \multicolumn{1}{l}{$p$} &
  \multicolumn{1}{c}{sym} &
  \multicolumn{1}{c}{$L$} &
  \multicolumn{1}{c}{Re~$\lambda_1$} &
  \multicolumn{1}{c}{Im~$\lambda_1$} &
  \multicolumn{1}{c}{Re~$\lambda_2$} &
  \multicolumn{1}{c}{Im~$\lambda_2$} \\
\hline
 0   & $\sigma_d$ &  4.000000 &  9.89898 &  0.00000 &  9.89898 &  0.00000\\
 1   & $C_3$      &  4.267949 & -11.7715 &  0.00000 &  9.28460 &  0.00000\\
 2   & $S_4$      &  4.296322 & -4.52562 &  9.49950 & -4.52562 & -9.49950\\
 01  & $\sigma_d$ &  8.316529 & -124.095 &  0.00000 &  88.4166 &  0.00000\\
 02  & $C_3$      &  8.320300 & -37.1479 &  98.0419 & -37.1479 & -98.0419\\
 12  & $S_4$      &  8.567170 &  117.644 &  0.00000 & -102.992 &  0.00000\\
 001 & $C_3$      & 12.321747 & -1240.54 &  0.00000 &  868.915 &  0.00000\\
 002 & $S_4$      & 12.322138 & -353.853 &  976.176 & -353.853 & -976.176\\
 011 & $\sigma_d$ & 12.580808 &  1449.55 &  0.00000 &  824.981 &  0.00000\\
 012 & $C_2$      & 12.617350 &  1192.83 &  0.00000 & -1020.66 &  0.00000\\
 021 & $C_3$      & 12.584068 &  1201.43 &  0.00000 & -996.800 &  0.00000\\
 022 & $S_4$      & 12.619948 & -755.582 &  804.976 & -755.582 & -804.976\\
 112 & $\sigma_d$ & 12.835715 & -496.339 &  1038.46 & -496.339 & -1038.46\\
 122 & $C_3$      & 12.863793 & -1100.56 &  0.00000 &  1219.28 &  0.00000
\end{tabular}
\end{center}
\end{table}

The semiclassical resonances of the four-sphere system are given by the poles 
of the function $g(k)$.
However, it is well known that the periodic orbit sum (\ref{gk}) does not 
converge in those regions where the physical poles are located.
For the three-disk system the cycle-expansion method \cite{Cvi89,Eck95,Wir99}
and harmonic inversion techniques \cite{Mai97c,Mai99} have proven to be  
powerful approaches for overcoming the convergence problems of the periodic 
orbit sum, and both methods can also be successfully applied to the 
four-sphere system.

The idea of the cycle-expansion method is to expand the Gutzwiller-Voros 
zeta function \cite{Gut90,Vor88}
\begin{equation}
 Z_{\rm GV}(k;z) = \exp\left\{-\sum_p\sum_{r=1}^\infty\frac{1}{r}
  \frac{(-z)^{rn_p}{\rm e}^{{\rm i}rkL_p}}
       {\sqrt{|\det({\mathbf M}_p^r-{\mathbf 1})|}}\right\}
\label{Z_GV}
\end{equation}
as a truncated power series in $z$, where $z$ is a book-keeping variable
which must be set to $z=1$.
The semiclassical resonances are obtained as the zeros of the cycle-expanded
zeta function (\ref{Z_GV}).

The harmonic inversion method is briefly explained as follows.
The Fourier transform of the function $g(k)$ in Eq.~(\ref{gk}) yields
the semiclassical signal
\begin{equation}
 C^{\rm sc}(L) = \sum_p \sum_{r=1}^\infty
   \frac{(-1)^{rn_p} L_p {\rm e}^{{\rm i}rkL_p}}
        {\sqrt{|\det({\mathbf M}_p^r-{\mathbf 1})|}} \delta(L-rL_p) \; ,
\label{C_sc}
\end{equation}
as a sum of $\delta$ functions.
The central idea of semiclassical quantization by harmonic inversion is
to adjust the semiclassical signal $C^{\rm sc}(L)$ with finite length
$L\le L_{\rm max}$ to its quantum mechanical analogue
\begin{equation}
 C^{\rm qm}(L) = \sum_n d_n {\rm e}^{-{\rm i}k_nL} \; ,
\label{C_qm}
\end{equation}
where the amplitudes $d_n$ and the semiclassical eigenvalues $k_n$ 
are free adjustable complex parameters.
Numerical recipes for extracting the parameters $\{d_n,k_n\}$ by harmonic 
inversion of the $\delta$ function signal (\ref{C_sc}) are given 
in \cite{Mai00,Bar01}.

\begin{figure}
\includegraphics[width=0.95\columnwidth]{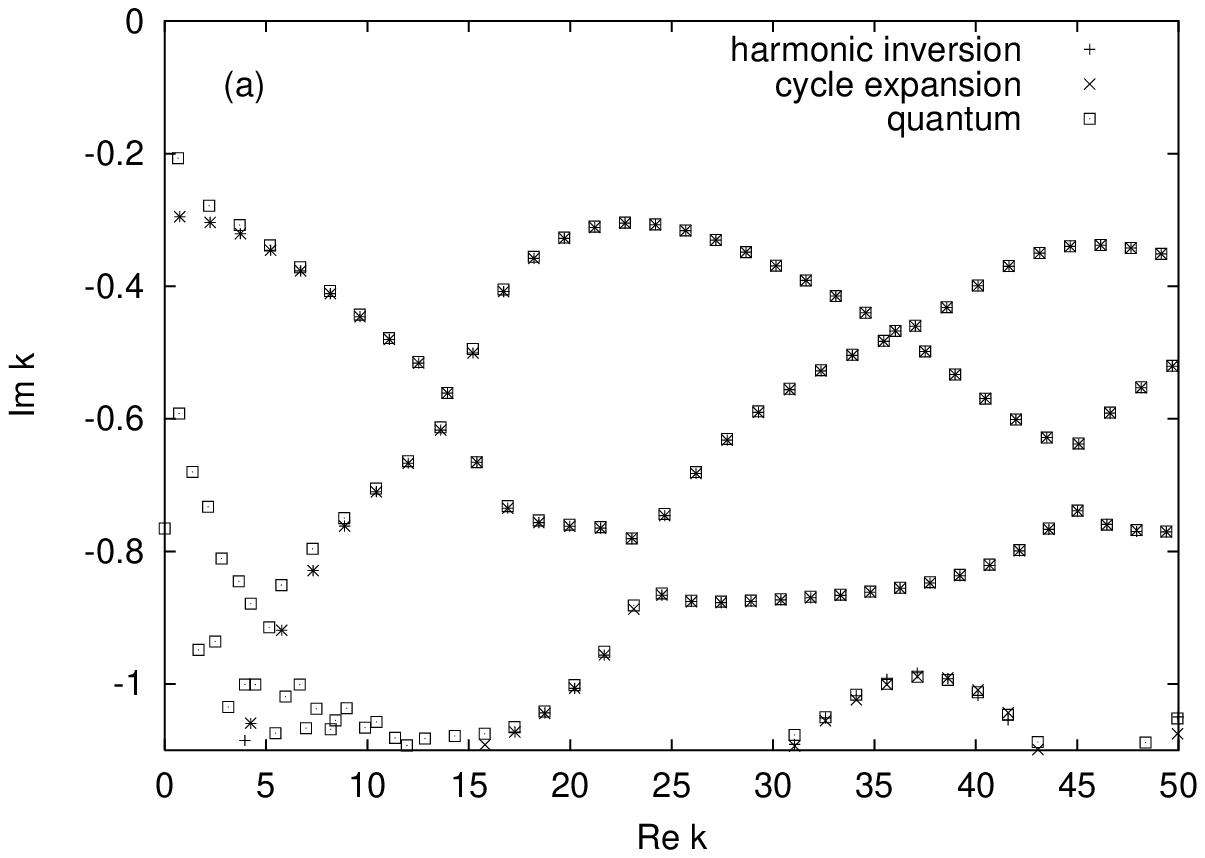}
\includegraphics[width=0.95\columnwidth]{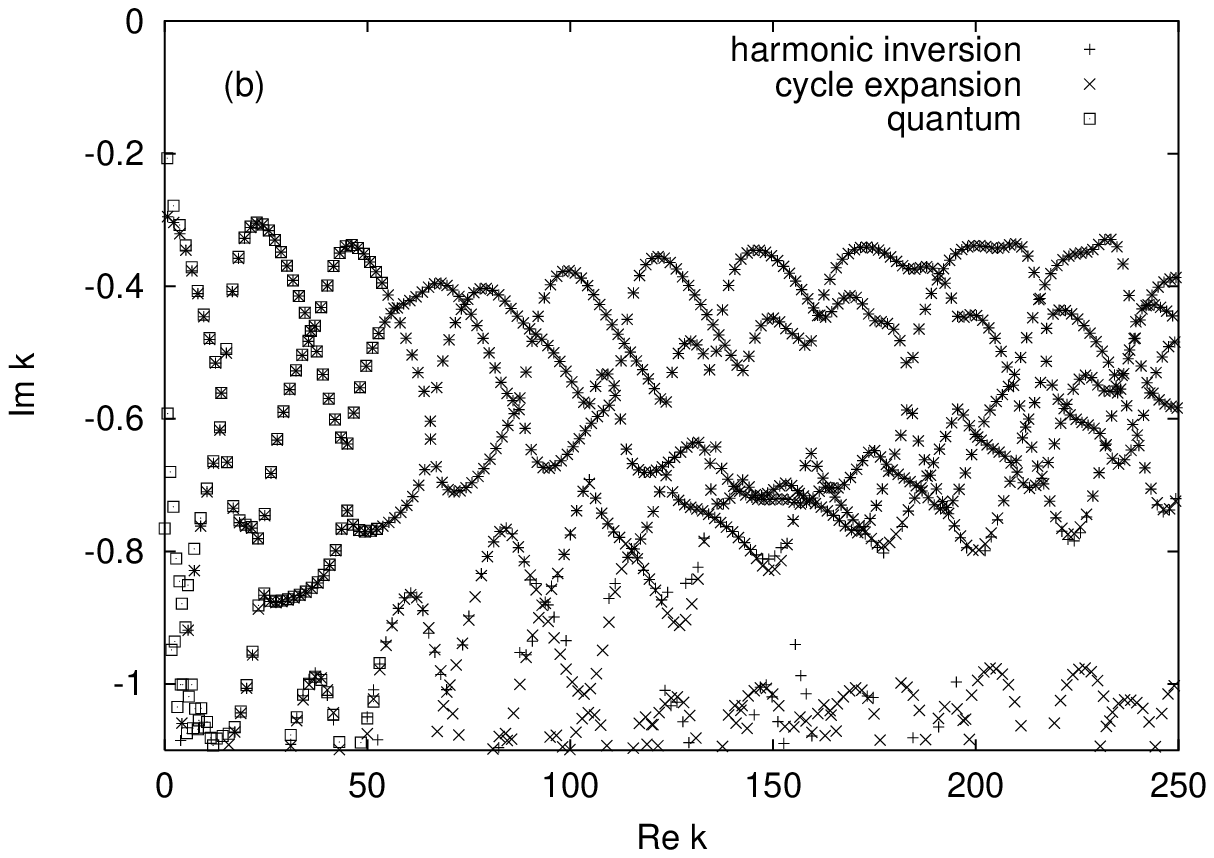}
\caption{$A_1$-resonances in the complex $k$-plane of the four-sphere system
with radius $a=1$ and center-to-center separation $R=6$. 
Squares: Quantum computations. 
Crosses and plus symbols: Semiclassical resonances obtained by cycle-expansion
and harmonic inversion methods, respectively.}
\label{fig1}
\end{figure}

The quantum mechanical and semiclassical $A_1$-resonances of the four-sphere
system with radius $a=1$ and center-to-center separation $R=6$ are presented
in Fig.~\ref{fig1}.
The quantum resonances marked by the squares have been obtained by solving
Eq.~(\ref{M_4sphere}) with matrices ${\mathbf M}_{lm,l'm'}$ of dimension
up to $(1134\times 1134)$, which is sufficient only to obtain converged
results in the region ${\rm Re}~k\lesssim 50$ (see Fig.~\ref{fig1}a).
By contrast, the semiclassical resonances can easily be obtained in a much
larger region, e.g., ${\rm Re}~k\le 250$ shown in Fig.~\ref{fig1}b.
The crosses mark the zeros of the cycle-expanded Gutzwiller-Voros 
zeta function (\ref{Z_GV}).
The cycle-expansion has been truncated at cycle length $n_{\rm max}=7$,
which means that a total set of just 508 primitive periodic orbits are
included in the calculation.
The plus symbols mark the semiclassical resonances obtained by harmonic
inversion of the periodic orbit signal (\ref{C_sc}) with signal length
$L_{\rm max}=60$ constructed from the set of 533830 primitive periodic orbits
with cycle length $n_p\le 14$.

In the region ${\rm Re}~k\le 50$ (Fig.~\ref{fig1}a) the quantum and
semiclassical resonances agree very well with a few exceptions.
The first few quantum resonances in the uppermost resonance band are narrower,
i.e., closer to the real axis than the corresponding semiclassical resonances.
A similar discrepancy between quantum and semiclassical resonances has already
been observed in the three-disk system \cite{Eck95,Wir99}.
Furthermore, in the region ${\rm Re}~k<15$ and ${\rm Im}~k<-0.5$ several
quantum resonances have been found (see the squares in Fig.~\ref{fig1}a),
which seem not to have any semiclassical analogue.
These resonances are related to the diffraction of waves at the spheres, and
its semiclassical description requires an extension of Gutzwiller's trace
formula and the inclusion of diffractive periodic orbits \cite{Vat94,Ros96}.
The semiclassical resonances obtained by either harmonic inversion or the
cycle-expansion method (the plus symbols and crosses in Fig.~\ref{fig1}b,
respectively) are generally in perfect agreement, except for the very 
broad resonances that lie deep in the complex plane, i.e., in the 
region ${\rm Im}~k\lesssim -0.8$.

For the four-sphere system with large separation $R=6a$ between the spheres
the cycle-expansion method is most efficient for the calculation of a large
number of resonances.
The reason is that the assumption of the cycle-expansion that the 
contributions of longer periodic orbits in the expansion of the
Gutz\-willer-Voros zeta function (\ref{Z_GV}) are shadowed by pseudo-orbits
composed of shorter periodic orbits is very well fulfilled.
The harmonic inversion method also allows for the calculation of a large
number of resonances, but requires a larger input set of periodic orbits.
In contrast, the quantum computations for this three-dimensional system
are very inefficient due to an unfavorable scaling of the dimension
of the matrix ${\mathbf M}_{lm,l'm'}$ in Eq.~(\ref{M_4sphere}).
For this reason, the quantum computations presented here have been 
restricted to the region ${\rm Re}~k\le 50$.
Of course, a more efficient quantum method for the four-sphere system than 
that of Ref.~\cite{Hen97} may in principle exist.
However, to the best of our knowledge no such method has been proposed 
in the literature to date.

\section{Summary and outlook}
In summary, we have applied Gutzwiller's periodic orbit theory to a system
with three degrees of freedom, viz.\ the four-sphere scattering problem.
For the first time, semiclassical resonances have been obtained for a
billiard system with genuinely three-dimensional periodic orbits.
For this system we have discussed the scaling properties of both quantum 
and semiclassical calculations and have demonstrated the superiority of 
semiclassical methods over quantum computations, i.e., semiclassical 
resonances could easily be obtained in energy regions which are unattainable 
with the established quantum method.
These results may encourage the investigation of other systems with three 
or more degrees of freedom with the goal to develop powerful semiclassical 
techniques, which are competitive with or even superior to quantum 
computations for a large variety of systems.

Interesting future work will also be the investigation of the four-sphere 
system when the spheres are moved towards each other.
In particular, the case of touching spheres with $R=2a$ is a real challenge,
because the cycle-expansion does not converge any more.
For touching spheres the symbolic dynamics is pruned in a similar way as
in the three-disk problem \cite{Han93}.
However, contrary to the closed three-disk system \cite{Tan91,Wei02}, the
four touching spheres do not form a bound system which means that the method
of Ref.~\cite{Tan91} combining the cycle-expansion method with a functional
equation cannot be applied.

\section*{Acknowledgements}
This work was supported by the National Science Foundation and the
Deut\-scher Akademischer Austauschdienst.~
P.A.D.\ thanks the Engineering and Physical Sciences Research Council
(UK) for support under grant number GR/N19519/01.
E.A.\ thanks J.M.\ and G.W.\ for the kind hospitality at the 
Institut f\"ur Theoretische Physik during his stay in Stuttgart.


\end{document}